\documentclass[10pt, twocolumn, notitlepage, superscriptaddress, prb, longbibliography]{revtex4-2}

\usepackage{here}
\usepackage{amsmath}         
\usepackage{graphicx}
\usepackage{braket}         
\usepackage{lettrine}
\usepackage{xcolor}
\usepackage{times}
\usepackage{hhline}
\usepackage{tabularx}
\usepackage{calc}
\usepackage{soul}
\usepackage[utf8]{inputenc}

\usepackage[colorlinks=true,linkcolor=blue, citecolor=blue, urlcolor=blue]{hyperref}%

\begin{document}

\title{Local and Global Reciprocity in Orbital-Charge-Coupled Transport}

\author{Dongwook~Go}
\email{d.go@fz-juelich.de}
\affiliation{Peter Gr\"unberg Institut, Forschungszentrum J\"ulich and JARA, 52425 J\"ulich, Germany}
\affiliation{Institute of Physics, Johannes Gutenberg University Mainz, 55099 Mainz, Germany}

\author{Tom~S.~Seifert}
\affiliation{Department of Physics, Freie Universit\"at Berlin, Berlin, Germany}

\author{Tobias~Kampfrath}
\affiliation{Department of Physics, Freie Universit\"at Berlin, Berlin, Germany}

\author{Kazuya~Ando}
\affiliation{Department of Applied Physics and Physico-Informatics, Keio University, Yokohama 223-8522, Japan}
\affiliation{Keio Institute of Pure and Applied Sciences (KiPAS), Keio University, Yokohama 223-8522, Japan}
\affiliation{Center for Spintronics Research Network (CSRN), Keio University, Yokohama 223-8522, Japan}


\author{Hyun-Woo~Lee
}
\affiliation{Department of Physics, Pohang University of Science and Technology, Pohang 37673, Korea}

\author{Yuriy~Mokrousov}
\affiliation{Peter Gr\"unberg Institut, Forschungszentrum J\"ulich and JARA, 52425 J\"ulich, Germany}
\affiliation{Institute of Physics, Johannes Gutenberg University Mainz, 55099 Mainz, Germany}

\begin{abstract}
The coupled transport of charge and orbital angular momentum (OAM) lies at the core of orbitronics. Here, we examine the reciprocal relation in orbital-charge-coupled transport in thin films, treating bulk and surface contributions on equal footing. We argue that the conventional definition of orbital current is ill-defiled, as it violates reciprocity due to the nonconservation of OAM. This issue is resolved by adopting the so-called \emph{proper} orbital current, which is directly linked to orbital accumulation. We establish the reciprocal relation for the \emph{global} (spatially integrated) response between orbital and charge currents, while showing that their \emph{local} (spatially resolved) responses can differ significantly. In particular, we find large surface contributions that may lead to nonreciprocity when currents are measured locally. These findings are supported by first-principles calculations on W(110) and Pt(111) thin films. In W(110), orbital-charge interconversion is strongly nonreciprocal at the layer level, despite exact reciprocity in the integrated response. Interestingly, spin-charge interconversion in W(110) remains nearly reciprocal even locally. In contrast, Pt(111) exhibits local nonreciprocity for both orbital-charge and spin-charge conversions, which we attribute to strong spin-orbit coupling. We propose that such local distinctions can be exploited to experimentally differentiate spin and orbital currents.
\end{abstract}

\date{\today}                 
\maketitle	      

\emph{Orbitronics} explores the interplay between charge and orbital degrees of freedom in nonequilibrium and steady-state transport~\cite{Go2021orbitronics}. A key phenomenon is the orbital Hall effect (OHE), the flow of electrons with finite orbital angular momentum (OAM) induced by an external electric field. The OHE has been theoretically predicted in hole-doped Si~\cite{Bernevig2005}, heavy~\cite{Tanaka2008, Kontani2009} and light~\cite{Jo2018, Go2018, Salemi2022, Go2024} transition metals, and two-dimensional materials~\cite{Bhowal2020, Canonico2020, Bhowal2021, Cysne2021}. Experiments have recently observed OHE-driven orbital accumulation in Ti~\cite{Choi2023} and Cr~\cite{Lyalin2023} thin films using the magneto-optical Kerr effect. In systems with broken inversion symmetry, orbital accumulation can also arise from an electric current by the orbital Edelstein effect (OEE), as shown in LaAlO$_3$/SrTiO$_3$ interfaces~\cite{ElHamdi2023, Johansson2021}, surface-oxidized Cu films~\cite{Ding2020, Go2021}, and chiral crystals~\cite{Yoda2018,Goebel2025}. Electrically induced OAM via the OHE or OEE has been detected in magnetotransport and torque measurements through its coupling to magnetization~\cite{Ding2020, Kim2021, Lee2021a, Lee2021b, Sala2022, Liao2022, Hayashi2023observation, Bose2023, Kim2023, Sala2023, Hayashi2023orbital, Taniguchi2023, Ding2024, Tang2024}. These orbitronic effects are not only fundamentally intriguing but also hold promise for applications. For example, nonequilibrium OAM can drive magnetization dynamics in spintronic devices~\cite{Go2020a, Go2020b, Go2023}, offering potentially greater efficiency than spin alone as a source of angular momentum.

While early studies in orbitronics mainly explored the electric response of OAM and its current, the reciprocal phenomena, where a charge current is driven by an \emph{orbital voltage}, i.e., a chemical potential difference depending on the electron's OAM (defined below), have only recently gained experimental attention~\cite{Santos2023, Xu2023, Wang2023, Seifert2023, ElHamdi2023, Hayashi2024,Kashiki2025}. These experiments typically involve two key microscopic processes. First, nonequilibrium OAM is generated by an external perturbation such as ferromagnetic resonance~\cite{Santos2023, ElHamdi2023, Hayashi2024}, the Seebeck effect~\cite{Santos2023, ElHamdi2023}, or optical excitation~\cite{Xu2023, Wang2023, Seifert2023}. Second, the resulting gradient of orbital voltage drives a charge current via orbital-to-charge conversion mechanisms like the inverse OHE or inverse OEE. While the reciprocity between orbital torque and pumping (in the first step) has been theoretically established~\cite{Go2025, Pezo2025, Han2025}, a formal framework and clear conceptual understanding of orbital-to-charge conversion remain lacking. 

Recent experiments hint at possible nonreciprocity in orbital-charge transport. For instance, THz spectroscopy on W thin films suggests a significant role of the inverse OEE at the surface~\cite{Seifert2023}, seemingly inconsistent with GHz torque measurements on similar samples~\cite{Hayashi2023observation}, which show signatures of bulk OHE. Furthermore, a recent thickness-dependent study via ferromagnetic resonance indicates that inverse effects may not be strictly reciprocal to their direct counterparts in each layer~\cite{Kashiki2025}. These discrepancies raise fundamental questions: could Onsager’s reciprocity~\cite{Onsager1931} be \emph{locally} violated, or do these observations still align with general principles such as the fluctuation-dissipation theorem~\cite{Kubo1966}?

In this Letter, we develop a formal theory of coupled transport between charge and orbital currents, focusing on their reciprocal relation. We explicitly consider finite-thickness films, treating bulk and surface contributions, i.e., direct and inverse OHEs and OEEs, on equal footing without assumptions on microscopic mechanisms. Crucially, we show that accounting for the nonconservation of orbital angular momentum (OAM) is essential to establish reciprocity in orbital-charge interconversion. To this end, we adopt the \emph{proper} orbital current, originally introduced by Shi \emph{et al.} for spin transport~\cite{Shi2006}, and rigorously prove the reciprocal relation for the \emph{global} (spatially integrated) responses for spin-to-charge and charge-to-spin inverconversions. Remarkably, the \emph{local} profiles of orbital and charge current responses in the charge-to-orbital and orbital-to-spin conversions, respectively, differ substantially, primarily due to large surface contributions, as demonstrated from first-principles for W(110) and Pt(111). However, in W(110), we do not observe such locally nonreciprocal behavior in spin-charge coupled transport. In Pt(111), both orbital-charge and spin-charge-coupled transports exhibit locally nonreciprocal behavior, which we attribute to strong spin-orbit coupling (SOC). These results suggest that spin and orbital transport may be experimentally distinguished via their differing local charge current profiles, which would resolve a key challenge in orbitronics.

In the following, we adopt phenomenological definitions of the direct OHE and inverse OHE, without referring to microscopic mechanisms in bulk and surfaces. That is, although the orbital-charge interconversion at surfaces is associated with the OEE, we simply denote it as the surface contribution to the OHE. Also, we use the atom-centered approximation for describing the local OAM near atomic nuclei~\cite{Hanke2016, Go2020b} rather than the Berry phase formalism~\cite{Thonhauser2005, Ceresoli2006, Ceresoli2010, Lopez2012, Xiao2005, Xiao2006, Shi2007} as the latter cannot be spatially resolved by its definition. The atom-centered approximation is known to give reasonable estimates of the OAM comparable to the experimental values for simple transition metals~\cite{Hanke2016}.

The commonly accepted definition of the direct OHE is an electric-field ($\boldsymbol{\mathcal{E}}$) induced orbital current, where the latter is defined as
\begin{eqnarray}
j_\alpha^{L_\gamma} = \frac{1}{2}\left( v_\alpha L_\gamma + L_\gamma v_\alpha \right),
\label{eq:conventional_current}
\end{eqnarray}
where $\boldsymbol{v}$ is the velocity and $\mathbf{L}$ is the OAM. In contrary to the direct OHE, the definition of the inverse OHE is yet to be agreed upon. The first problem one encounters is how to define the perturbation, i.e. what orbital voltage is, whose gradient generates orbital current.
As the electric voltage is the work required to displace electric charge under an electric field, the orbital voltage can be analogously defined as the work required to displace an OAM-polarized electron under an OAM-dependent electric field,
\begin{eqnarray}
V^L = - \mathcal{E}_\alpha^{L_\gamma} \mathcal{P}_\alpha^{L_\gamma}.
\label{eq:orbital_voltage}
\end{eqnarray}
Here, $\mathcal{E}_\alpha^{L_\gamma}$ is an $L_\gamma$-dependent electric field in $\alpha$ direction, which couples to \emph{orbital dipole} 
\begin{equation}
\mathcal{P}_\alpha^{L_\gamma} = \frac{1}{2}( r_\alpha L_\gamma + L_\gamma r_\alpha ).
\label{eq:orbital_dipole}
\end{equation}
For instance, $\mathcal{E}_\alpha^{L_\gamma}$ describes the potential gradient dependent on the expectation value of $L_\gamma$ for an electronic state, which tends to induce $L_\gamma$-polarized orbital current along $\alpha$ direction. Therefore, we can define the inverse OHE as the response of the charge current to the orbital voltage.

However, in the definitions of Eqs.~\eqref{eq:conventional_current}-\eqref{eq:orbital_dipole}, one notices that the conjugate relation is improper, that is, $j_\alpha^{L_\gamma} \neq {d\mathcal{P}_\alpha^{L_\gamma}}/{dt}$, unlike the relation between the charge dipole and current. This is the result of the nonconservation of the OAM in crystals, which persists regardless of the SOC. Therefore, the conventional orbital current fails to satisfy the Onsager's reciprocal relation.

Instead, if we define orbital current by 
\begin{gather}
\mathcal{J}_\alpha^{L_\gamma} = \frac{d\mathcal{P}_\alpha^{L_\gamma}}{dt}
=
j_\alpha^{L_\gamma}
+
\mathcal{P}_\alpha^{T_\gamma},
\label{eq:proper_current}
\end{gather}
the conjugation relation between orbital current and voltage is recovered. This definition is known as \emph{proper orbital current}, first proposed by Shi \emph{et al.} in the description of spin currents in the presence of the SOC~\cite{Shi2006}. Here, we denote the first term by \emph{conventional orbital current} [Eq.~\eqref{eq:conventional_current}] and the second term 
\begin{equation}
\label{eq:torque_dipole}
\mathcal{P}_\alpha^{T_{L_\gamma}}=\frac{1}{2}( r_\alpha {T}_{L_\gamma} + {T}_{L_\gamma} r_\alpha)
\end{equation}
by \emph{torque dipole}. Here, the torque on the OAM ${T}_{L_\gamma}={dL_\gamma}/{dt}= 
\left[
L_\gamma, \mathcal{H}
\right]/i\hbar
$
arises from the interaction between the OAM and the lattice in the electronic Hamiltonian $\mathcal{H}$. 

The proper orbital current satisfies the continuity equation, ${\partial L_\gamma}/{\partial t} + \partial_\alpha \mathcal{J}_\alpha^{L_\gamma}=0$ ~\cite{Shi2006, Go2020b}. We show that the intrinsic response of the proper orbital current is directly proportional to the orbital dipole (antisymmetric accumulation of the OAM at top and bottom surfaces) by
\begin{eqnarray}
\label{eq:correspondence}
\left\langle \mathcal{J}_{\alpha}^{L_\gamma} \right\rangle 
=
\frac{1}{\tau}
\left\langle \mathcal{P}_{\alpha}^{L_\gamma} \right\rangle,
\end{eqnarray}
where $\tau$ is the relaxation time. We prove this relation and numerically demonstrate it in the Supplemental Material~\cite{Supplemental}. Equation~\eqref{eq:correspondence} suggests that the OHE in the bulk and the OEE at surfaces are described in a unified manner in the formulation by the proper orbital current. We also note that this is equivalent to the continuity equation derived in Ref.~\cite{Go2020b}. Therefore, calculating an electrical response of the proper orbital current is equivalent to calculating the current-induced OAM accumulation.

Although the orbital and torque dipoles are physically meaningful only when integrated over space due to the ambiguity of the coordinate origin, as the electric polarization is~\cite{King-Smith1993, Vanderbilt1993, Resta1994}, their spatial profiles can still be useful, especially when the inversion symmetry is \emph{globally} present, as explained in Ref.~\cite{Shi2006}. This prevents any torque response in the bulk under the perturbation by an external electric field. Thus, in the direct OHE, the torque dipole contribution can be considered as a purely surface effect, which tends to grow linearly with the film thickness by its definition. In contrast, the conventional orbital current is dominated by the bulk contribution as the film thickness grows. Meanwhile, in the inverse OHE, the orbital voltage is applied globally, which is unambiguously defined, and spatially resolving the charge current response is also a well-defined procedure.

The conductivity tensors for the direct and inverse OHEs are defined by
\begin{subequations}
\begin{eqnarray}
\sigma_{\mathrm{dir},\alpha\beta}^{L_\gamma}
= \left\langle  \mathcal{J}_\alpha^{L_\gamma} \right\rangle / \mathcal{E}_\beta,
\\
\sigma_{\mathrm{inv},\alpha\beta}^{L_\gamma}
= \left\langle  j_\alpha^{-e} \right\rangle / \mathcal{E}_\beta^{L_\gamma},
\end{eqnarray}
\end{subequations}
where $\langle \mathcal{J}_\alpha^{L_\gamma} \rangle$ is the response of the proper orbital current to an external electric field, and $\langle j_\alpha^{-e} \rangle$ is the response of the charge current to an orbital voltage. Now that orbital voltage and orbital current are the proper conjugates, the reciprocal relation is satisfied,
\begin{eqnarray}
\label{eq:reciprocity_global}
\sigma_{\mathrm{dir},\alpha\beta}^{L_\gamma} = 
-\sigma_{\mathrm{inv},\beta\alpha}^{L_\gamma},
\end{eqnarray}
which is explicitly demonstrated for real materials in the below. However, we note that it is valid only for the \emph{global} (volume-integrated) responses.


\begin{figure*}[t!]
\centering
\includegraphics[angle=0, width=0.99\textwidth]{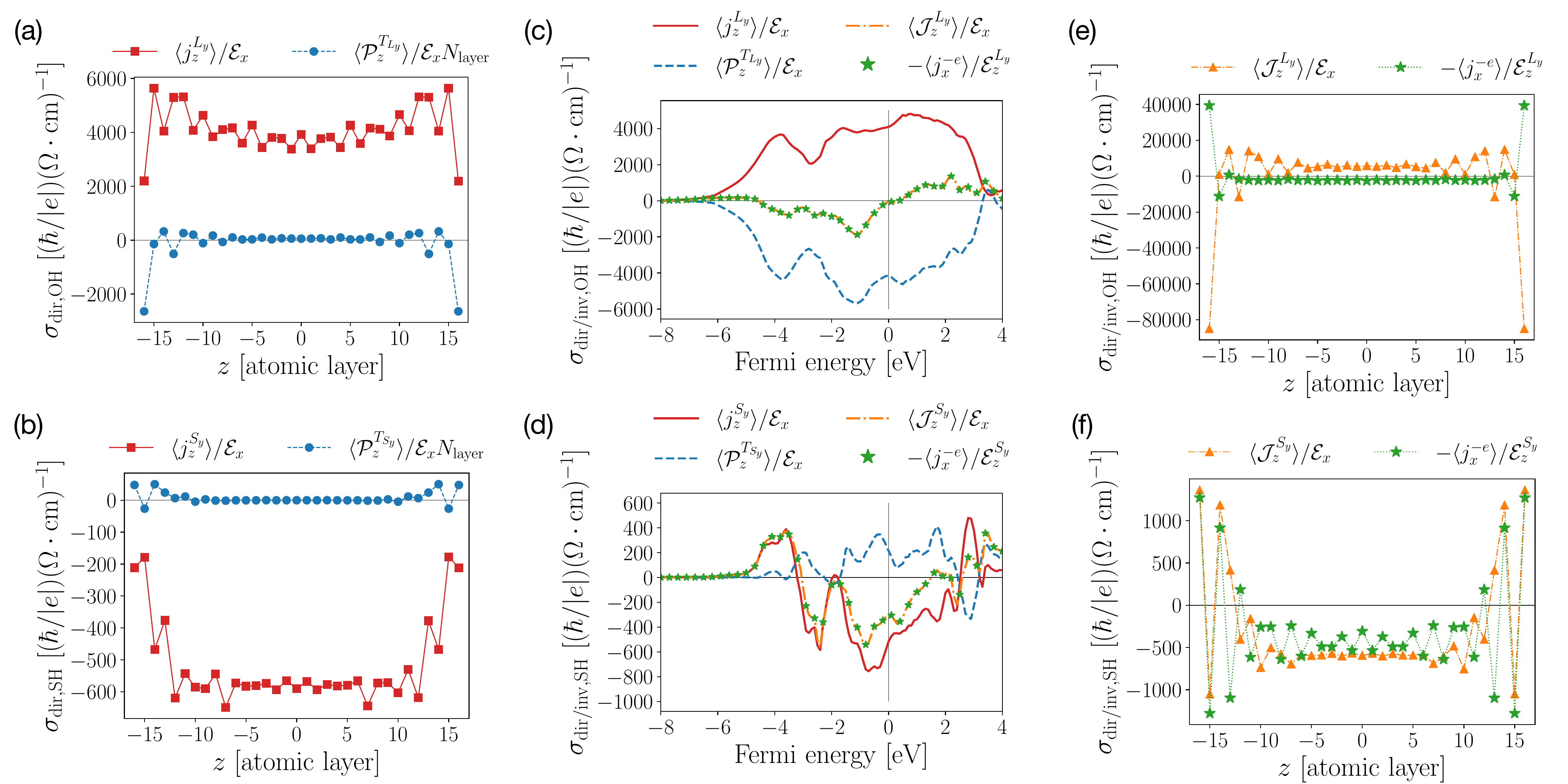}
\caption{
First-principles calculation of the direct and inverse OHE (a,c,e) and SHE (b,d,e) in a W(110) thin film. (a,b) Local electric response of conventional orbital/spin current (red square symbols) and torque dipole (blue circle symbols) by the direct OHE/SHE. (c,d) Fermi energy dependence of the global electric responses of conventional orbital/spin current (red solid lines) and torque dipole (blue dashed lines). The sum of conventional orbital/spin current and torque dipole, proper orbital/spin current, is shown in orange dash-dot lines, which is exactly reciprocal to the response of charge current by orbital/spin voltage (green star symbols). (e,f) Comparison of the local responses of proper orbital/spin current (orange triangle symbols) and charge current (green star symbols) in the direct and inverse OHE/SHE, respectively. The local responses are substantially different in the direct and inverse OHEs, while those for the SHEs are reciprocal even locally.
\label{fig:1}
}
\end{figure*}

We now proceed with the demonstration of the reciprocal relation in orbital-charge-coupled transport, by considering explicitly finite-thickness films to examine both bulk and surface contributions on an equal footing. Our findings are demonstrated for W and Pt thin films in bcc(110) and fcc(111) stacks. We primarily focus on the results for W in the main text, while the discussion of the Pt case can be found in the Supplemental Materials~\cite{Supplemental} together with details on the first-principles calculation. The W film is finite ($N_\mathrm{layer}=33$) in $z\parallel [110]$ direction and periodic in $x\parallel [\bar{1}10]$ and $y \parallel [001]$ directions. We set the coordinate origin at the center of films. The crystal momentum $\mathbf{k}$ is defined in the $xy$ plane.

For the direct OHE, we calculate the local response of the orbital current $\mathcal{J}_z^{L_y}$ induced by an external electric field $\mathcal{E}_x$ from the intrinsic Kubo formalism,
\begin{eqnarray}
\label{eq:OHE_direct}
& & 
\sigma_{\mathrm{dir},zx}^{L_y}(z)
=
\\
& & 
-\frac{e\hbar}{D}
\int \frac{d^2k}{(2\pi)^2}
\sum_{n\neq n'}
(f_{n\mathbf{k}}-f_{n'\mathbf{k}}) 
\sum_{z'}
\Omega_{zx,nn'}^{L_y}(\mathbf{k};z,z'),
\nonumber
\end{eqnarray}
where $\hbar$ is the reduced Planck constant, $D$ is the thickness of one atomic layer, and $f_{n\mathbf{k}}$ is the Fermi-Dirac distribution function for a Bloch state $\psi_{n\mathbf{k}}$ with its energy $E_{n\mathbf{k}}$. Here,
\begin{eqnarray}
\label{eq:orbital_Berry_direct}
& &
\Omega_{zx,nn'}^{L_y}(\mathbf{k};z, z')
\\
& &
=
\mathrm{Im}
\left[ 
\frac{
\bra{\psi_{n\mathbf{k}}}
\mathcal{J}_z^{L_y} (z) 
\ket{\psi_{n'\mathbf{k}}}
\bra{\psi_{n'\mathbf{k}}}
v_x (z')
\ket{\psi_{n\mathbf{k}}}
}{(E_{n\mathbf{k}}-E_{n'\mathbf{k}})^2 + \eta^2}
\right]
\nonumber 
\end{eqnarray}
is the correlation function between the orbital current at $z$ and the velocity at $z'$. On the other hand, in the inverse OHE, the local response of the charge current by orbital-dependent electric field $\mathcal{E}_z^{L_y}$ is given by
\begin{eqnarray}
\label{eq:OHE_inverse}
& &
\sigma_{\mathrm{inv},xz}^{L_y}(z)
=
\\
& &
-
\frac{e\hbar}{D}
\int \frac{d^2k}{(2\pi)^2}
\sum_{n\neq n'}
(f_{n\mathbf{k}}-f_{n'\mathbf{k}}) 
\sum_{z'}
\widetilde{\Omega}_{xz,nn'}^{L_y}(\mathbf{k};z,z'),
\nonumber
\end{eqnarray}
where
\begin{eqnarray}
& &
\widetilde{\Omega}_{xz,nn'}^{L_y}(\mathbf{k};z, z')
=
\label{eq:orbital_Berry_inverse}
\\
& &
\mathrm{Im}
\left[ 
\frac{
\bra{\psi_{n\mathbf{k}}}
v_x (z)
\ket{\psi_{n'\mathbf{k}}}
\bra{\psi_{n'\mathbf{k}}}
\mathcal{J}_z^{L_y} (z')
\ket{\psi_{n\mathbf{k}}}
}{(E_{n\mathbf{k}}-E_{n'\mathbf{k}})^2 + \eta^2}
\right]
\nonumber 
\end{eqnarray}
is the correlation function between the velocity at $z$ and the orbital current at $z'$. For an arbitrary operator $\mathcal{O}$, we define the local projection by $\mathcal{O}(z)=[\mathcal{O} P(z) + P(z) \mathcal{O}]/2$, where $P(z)$ is the projection operator on the layer at $z$~\cite{Go2020b}. Note that the global orbital Hall conductivities are given as the average over $z$, 
\begin{equation}
\sigma_\mathrm{dir/inv}^{\mathbf{L}}=
\frac{1}{N_\mathrm{layer}}\sum_z 
\sigma_\mathrm{dir/inv}^{\mathbf{L}} (z).
\end{equation}
In numerics, we set $\eta=0.1\ \mathrm{eV}$ for the broadening of the energy spectrum, which effectively captures disorder effects and removes quantum well oscillations, and $400\times 400$ $\mathbf{k}$-mesh is used for the integral over the two-dimensional first Brillouin zone.

In Fig.~\ref{fig:1}(a), we show the electric response of the proper orbital current separately for the conventional current and the torque dipole. Conventional current arises in the bulk as well as at the surfaces. On the other hand, torque dipole appears only at the surfaces, whose numerical value is divided by $N_\mathrm{layer}$ in the plot as it grows linearly with the system size by its definition. The absence of torque dipole in the bulk is due to the presence of inversion symmetry. We remark that a finite torque dipole reflects the non-conservation of the OAM. At the surface, the anisotropic crystal potential, which differs from the bulk crystal potential, efficiently mediates the angular momentum exchange between the electron and lattice. Meanwhile, the analogous plot for the direct SHE [Fig.~\ref{fig:1}(b)] shows negligible contribution of torque dipole compared to conventional current.

The global (volume-integrated) responses for the proper current and torque dipole are shown in Figs.~\ref{fig:1}(c) and \ref{fig:1}(d), respectively, for the direct OHE and SHE. For the direct OHE, we find a general tendency that the conventional orbital current cancels with the torque dipole. This means, despite the conventional current arising in the bulk, the torque dipole at a surface results in loss of angular momentum, and only the remaining part results in accumulation at the surface. Also, smooth variations in both conventional current and torque dipole over a wide energy range in Fig.~\ref{fig:1}(c) shows that the relevant interaction for orbital-charge-coupled transport, which is the crystal-field potential, is of the order of eV. On the other hand, for the direct SHE in Fig.~\ref{fig:1}(d), we observe rapidly varying peaks and dips at different energies, which reflects coincidental hotspots due to band crossings gapped by the SOC. As a result, the torque dipole response in the direct SHE is pronounced only at particular energies and is generally smaller than the conventional current. Thus, the total proper spin current is mostly dominated by the contribution of the conventional current. 

We remark that the two microscopic correlation functions, responsible for the direct and inverse OHEs [Eqs.~\eqref{eq:orbital_Berry_direct} and \eqref{eq:orbital_Berry_inverse}] are related by time-reversal,
\begin{equation}
\label{eq:microscopic_correlation_local}
\Omega_{zx,nn'}^{L_y}(\mathbf{k};z,z')
=
-\widetilde{\Omega}_{xz,nn'}^{L_y}(-\mathbf{k};z',z),
\end{equation}
which is the hallmark of the fluctuation-dissipation theorem, relating the microscopic fluctuations and the macroscopic responses~\cite{Onsager1931, Kubo1966}. Therefore, summing over $z$ and $z'$ and integrating over $\mathbf{k}$ results in the global reciprocity between the direct and inverse OHEs~[Eq.~\eqref{eq:reciprocity_global}]. In Fig.~\ref{fig:1}(c,d) we numerically demonstrate the global reciprocal relation for the OHE and SHE, respectively, where the global response of the inverse OHE and SHE are shown by green star symbols.

However, the direct and inverse OHEs, whose responses are locally defined at $z$ for a perturbation globally applied over a film, they may be locally nonreciprocal,
\begin{eqnarray}
\label{eq:reciprocity_local}
\sigma_{\mathrm{dir},zx}^{L_y}(z) 
\neq  
-\sigma_{\mathrm{inv},xz}^{L_y}(z).
\end{eqnarray}
Mathematically, this is because the reciprocal relation holds only if the layer indices $z$ and $z'$ are switched in the microscopic correlations, Eq.~\eqref{eq:microscopic_correlation_local}. Interestingly, we find that the local reciprocity between the direct and inverse OHEs is strongly violated, while the direct and inverse SHEs are nearly reciprocal locally at each $z$. Figure~\ref{fig:1}(e) clearly shows distinct local responses in the direct and inverse OHEs, whose sums correspond to the values at zero Fermi energy in Fig.~\ref{fig:1}(c). We find gigantic contributions at the surfaces, with unexpected signs according to the reciprocal relation. Moreover, even in the bulk, the signs of the orbital Hall angles are different between the direct and inverse OHEs. On the other hand, as shown in Fig.~\ref{fig:1}(f), the direct and inverse SHEs are nearly reciprocal even locally despite slight deviation, both in the bulk and at the surface. 

Considering that the SHEs are computed by using the same Kubo formula [Eqs.~\eqref{eq:OHE_direct}-\eqref{eq:orbital_Berry_inverse}] used for the OHEs, simply by replacing the OAM $\mathbf{L}$ with the spin $\mathbf{S}$, the different behaviors of the OHEs and SHEs reflect their distinct microscopic natures. While the spin is almost a good quantum number due to the Kramer's degeneracy even in the presence of the SOC, the OAM is not conserved at all, as it exchanges angular momentum with the lattice. By separating the perturbation in the inverse OHE by the conventional current and torque dipole contributions [Eq.~\eqref{eq:orbital_Berry_inverse}] in our numerical calculation, we find that the the `anomalous signs' of the inverse OHE, both in the bulk and at surfaces, originate from the torque dipole, e.g. nonconservation of the OAM~\cite{Supplemental}.

\newcolumntype{M}[1]{>{\centering\arraybackslash}m{#1}}
\renewcommand{\arraystretch}{1.2}
\begin{table*}[t!]
\centering
\begin{tabular}{| M{2.2cm} | M{1.7cm} | M{1.7cm} | M{1.7cm} || M{2.2cm} | M{1.7cm} | M{1.7cm} | M{1.7cm} |}
\hline \hline  
W(110) & Bulk & Surface & Total & Pt(111) & Bulk & Surface & Total
\\
\hline \hline 
\ \ \ $\sigma_{\mathrm{dir},zx}^{L_y}$ & $+4571$ & $-4626$ & $-55$ & \ \ \ $\sigma_{\mathrm{dir},zx}^{L_y}$ & $+1895$ & $-1256$ & $+639$ 
\\
\hline 
$-\sigma_{\mathrm{inv},xz}^{L_y}$ & $-1891$ & $+1836$ & $-55$ & $-\sigma_{\mathrm{inv},xz}^{L_y}$ & $+1833$ & $-1244$ & $+639$ 
\\
\hline
\ \ \ $\sigma_{\mathrm{dir},zx}^{S_y}$ & $-300$ & $-2$ & $-302$ & \ \ \ $\sigma_{\mathrm{dir},zx}^{S_y}$ & $+1244$ & $-559$ & $+685$
\\
\hline 
$-\sigma_{\mathrm{inv},xz}^{S_y}$ & $-299$ & $-3$ & $-302$ & $-\sigma_{\mathrm{inv},xz}^{S_y}$ & $+457$ & $+228$ & $+685$
\\
\hline  
$-\sigma_{\mathrm{dir},zy}^{L_x}$ & $+5773$ & $-5733$ & $+40$ & $-\sigma_{\mathrm{dir},zy}^{L_x}$ & $+1865$ & $-1215$ & $+650$
\\
\hline 
\ \ \ $\sigma_{\mathrm{inv},yz}^{L_x}$ & $-1591$ & $+1631$ & $+40$ & \ \ \ $\sigma_{\mathrm{inv},yz}^{L_x}$ & $+1923$ & $-1274$ & $+650$
\\
\hline
$-\sigma_{\mathrm{dir},zy}^{S_x}$ & $-341$ & $-77$ & $-418$ & $-\sigma_{\mathrm{dir},zy}^{S_x}$ & $+1244$ & $-551$ & $+693$
\\
\hline 
\ \ \ $\sigma_{\mathrm{inv},yz}^{S_x}$ & $-443$ & $+25$ & $-418$ & \ \ \ $\sigma_{\mathrm{inv},yz}^{S_x}$ & $+469$ & $+223$ & $+693$
\\
\hline \hline 
\end{tabular}
\caption{
Decomposition of the surface and bulk contributions to the direct and inverse OHE and SHE in W(110) and Pt(111) thin films, in unit of $(e/\hbar)(\Omega\cdot\mathrm{cm})^{-1}$. The surface is defined as 3 atomic layers from the vacuum, while the rest is considered the bulk. 
}
\label{tab:1} 
\end{table*}

Table~\ref{tab:1} summarizes the surface and bulk contributions to the direct and inverse OHEs and SHEs in W(110) and Pt(111) films. Detailed results including their spatial profiles are shown in the Supplemental Material~\cite{Supplemental}. For W(110), we also show the results for $\boldsymbol{\mathcal{E}}\parallel \hat{\mathbf{y}}$, whose main features such as the sign in the bulk and surface agree with the case for $\boldsymbol{\mathcal{E}}\parallel \hat{\mathbf{x}}$. Meanwhile, in Pt, the local responses of the direct and inverse OHEs/SHEs are different both in the bulk and at the surface. At the Fermi energy, the OHEs seem more reciprocal locally than the SHEs are. We attribute this behavior to the pronounced SOC in Pt. In line with this, the behavior is drastically  different over a wide range of energies away from the actual band filling. Here, the SHEs are mostly reciprocal even locally, but the direct and inverse OHEs exhibit substantially different local responses, similar to the case of W.

We propose that the distinct local responses in the direct and inverse OHEs can be a smoking gun feature of orbital current, which can be used to distinguish it from spin current in experiment. For example, we believe our prediction of a positive orbital Hall angle of the inverse OHE at W surfaces [Fig.~\ref{fig:1}(e)] explains the THz spectroscopy experiment~\cite{Seifert2023}. On the other hand, the current-induced orbital torque measurement~\cite{Hayashi2023observation} clearly shows the bulk origin of the direct OHE, whose sign is positive, as also captured by our calculation [Fig.~\ref{fig:1}(e)]. We remark that a recent experiment on the thickness-dependence of the direct and inverse OHEs in ferromagnetic resonance in W and Pt films~\cite{Kashiki2025} strongly suggests the breakdown of local reciprocity between the direct and inverse OHEs, which is in agreement with our theoretical predictions shown in Table~\ref{tab:1}, both qualitatively and quantitatively. 


We note that the surface contribution in both direct and inverse OHEs is sensitive to the boundary condition, while the bulk contribution is robust. That is, at different surfaces, e.g. interfaced with a substrate, capping, or magnetic layer, the surface contributions may become different. This suggests that the proper orbital/spin current in the bulk described by $\mathbf{k}$-space formalisms~\cite{Shi2006, Liu2023, Tamaya2024} needs to be augmented by the surface contributions in open boundary condition.

The proper spin current was shown to be crucial in quantitative prediction of the SHE~\cite{Zhang2008}. More recently, Xiao and Niu applied a similar idea on Bogoliubov quasiparticles, where charge is not conserved~\cite{Cong2021}. In orbitronics, Liu \emph{et al.} considered the proper orbital current, beyond the atom-centered approximation on the OAM~\cite{Liu2025}. Investigation on the reciprocal transport in these objects remains as an appealing direction of research in the near future.

In summary, we have developed a theory of the reciprocal transport between orbital and charge currents by adopting the notion of proper orbital current. This takes the nonconservation of the OAM into account, which is consistent with the definition of orbital voltage. We have shown that the local responses of orbital current and charge current may be significantly different, in the direct and inverse OHEs, respectively, although the global responses are completely reciprocal. On the other hand, we find that this feature is not particularly pronounced between the direct and inverse SHEs. Therefore, we propose that breakdown of the local reciprocity can be used to distinguish orbital current from spin current, for which experimental investigation is encouraged.

D.G. acknowledges inspiring discussion with Aur\'elien Manchon, Daegeun Jo, Giovanni Vignale, and Qian Niu on the proper definition of orbital currents. We gratefully acknowledge the J\"ulich Supercomputing Centre for providing computational resources under project jiff40. This work was funded by the EIC Pathfinder OPEN grant 101129641 ``OBELIX'' and by the Deutsche Forschungsgemeinschaft (DFG, German Research Foundation) $-$ TRR 173/3 $-$ 268565370 (project A11) and TRR 288 $-$ 422213477 (project B06). K.A. acknowledges the support by JSPS KAKENHI (Grant Number 22H04964) and MEXT Initiative to Establish Next-generation Novel Integrated Circuits Centers (X-NICS) (Grant Number JPJ011438). T.S.S. and T.K. acknowledge funding by the German Research Foundation through the collaborative research center SFB TRR 227 ``Ultrafast spin dynamics'' (project ID 328545488, projects A05 and B02) and the priority program SPP2314 ``INTEREST'' (project ID GE 3288 2-1, project ITISA), the Federal Ministry of Education and Research (BMBF), and the European Research Council through ERC-2023-AdG ORBITERA (grant No. 101142285). H.W.L. acknowledges the funding by the National Research Foundation of Korea (NRF) (Grant Number RS-2024-00410027).

\let\oldaddcontentsline\addcontentsline
\renewcommand{\addcontentsline}[3]{}

\newpage 

\clearpage 
\widetext

\setcounter{equation}{0}
\setcounter{figure}{0}
\setcounter{table}{0}
\setcounter{page}{1}

\renewcommand{\theequation}{S\arabic{equation}}
\renewcommand{\thefigure}{S\arabic{figure}}
\renewcommand{\bibnumfmt}[1]{[S#1]}
\renewcommand{\citenumfont}[1]{S#1}
\renewcommand{\thepage}{S\arabic{page}}

\makeatletter
\def\@hangfrom@section#1#2#3{\@hangfrom{#1#2}#3}
\def\@hangfroms@section#1#2{#1#2}
\makeatother

\makeatletter
\renewcommand*{\thesection}{\arabic{section}}
\renewcommand*{\thesubsection}{\thesection.\arabic{subsection}}
\renewcommand*{\p@subsection}{}
\renewcommand*{\thesubsubsection}{\thesubsection.\arabic{subsubsection}}
\renewcommand*{\p@subsubsection}{}
\makeatother

\linespread{1.0}

\begin{center}
\Large \bf    
Supplemental Material for
\\
``Local and Global Reciprocity in Orbital-Charge-Coupled Transport''
\end{center}

\begin{center}
Dongwook Go, Tom~S.~Seifert, Tobias Kampfrath, Kazuya Ando, Hyun-Woo Lee, and Yuriy Mokrousov
\end{center}

\let\addcontentsline\oldaddcontentsline
\tableofcontents

\section{Details on Computation Methods}

We carry out first-principles calculation of the direct and inverse orbital/spin Hall effects by the Wannier interpolation technique, which is generally a three-step procedure: (1) Self-consistent density functional theory calculation, (2) Construction of maximally localized Wannier functions, and (3) Evaluation of response functions for the direct and inverse orbital/spin Hall effects from the Wannier representation. On (1), we use the film mode of \texttt{FLEUR} code~\cite{fleur}, which implements the full-potential linearly augmented plane wave method of the density functional theory~\cite{Wimmer1981}. For exchange and correlation effects, we use Perdew-Burke-Ernzerhof functional i nthe scheme of the generalized gradient approximation~\cite{Perdew1996}. On (2) we use \texttt{WANNIER90} code~\cite{Pizzi2020} which is interfaced with \texttt{FLEUR} code~\cite{Freimuth2008}. On (3), we use a custom-built code, which has proven its accuracy in our previous works, for example, in Refs.~\cite{Go2020b, Go2024}. For both W(110) and Pt(111), the films are composed of 33 atomic layers and are constructed assuming the same atomic coordinates as those for the bulk bcc and fcc crystals, respectively, without performing structural relaxation. Detailed computation parameters are summarized in Tab.~\ref{tab:params}.

\newcolumntype{M}[1]{>{\centering\arraybackslash}m{#1}}
\renewcommand{\arraystretch}{1.2}
\begin{table*}[h!]
\begin{center}
\centering
\begin{tabular}{M{6.0cm} | M{5.0cm} | M{5.0cm}}
\hline \hline
& W(110) & Pt(111) \\ \hline \hline 
Bulk structure and lattice constant & bcc, $5.96 a_0$ & fcc, $7.41 a_0$ \\ \hline
$D_\mathrm{vac}$, $\widetilde{D}$  & $140.43 a_0$, $144.11 a_0$ & $142.44 a_0$, $146.17 a_0$ \\ \hline
$R_\mathrm{MT}$, $l_\mathrm{max}$ & $2.5 a_0$, 12 & $2.5 a_0$, 12 \\ \hline 
$K_\mathrm{max}$, $G_\mathrm{max}$, $G_\mathrm{max, XC}$ & $4.0a_0^{-1}$, $12.2a_0^{-1}$, $10.1 a_0^{-1}$ & $4.0a_0^{-1}$, $12.0a_0^{-1}$, $10.0 a_0^{-1}$ \\ \hline
ab-initio $\mathbf{k}$-mesh & $16\times 16$ & $16\times 16$ \\ \hline 
Wannier initial projections & $spd$ on each atomic center & $spd$ on each atomic center \\ \hline
Maximum frozen energy window & $+2.0\ \mathrm{eV}$ above the Fermi energy & $+3.0\ \mathrm{eV}$ above the Fermi energy \\ \hline
Interpolation $\mathbf{k}$-mesh & $400\times 400$ & $400\times 400$ \\ \hline
\hline
\end{tabular}
\caption{
\label{tab:params}
Summary of computation parameters.
}
\end{center}
\end{table*}

\section{Correspondence between Proper Orbital Current and Orbital Accumulation}

Here, we prove Eq.~(6) in the main text, the relation between intrinsic response of proper orbital current and extrinsic response of orbital accumulation in the steady state under a dc external electric field. The proof follows from \emph{the interband-intraband correspondence} in Kubo linear response, which is shown in Ref.~\cite{Go2020b}. This is summarized by the following relation:
\begin{equation}
\frac{1}{\tau}
\left\langle 
\mathcal{O}
\right\rangle^\mathrm{intra}
=
\left\langle 
\frac{d\mathcal{O}}{dt}
\right\rangle^\mathrm{inter},
\label{eq:correspondence}
\end{equation}
where $\mathcal{O}$ is an arbitrary local observable, its time-derivative is defined through the Heisenberg's equation of motion, ${d\mathcal{O}}/{dt}=[\mathcal{O},\mathcal{H}]/i\hbar$, and the superscripts `intra' and `inter' indicate extrinsic and intrinsic parts of the responses under relaxation time ($\tau$) approximation, respectively. The intraband contribution is driven by the shift of occupation functions due to momentum relaxation:
\begin{equation}
\left\langle 
\mathcal{O}
\right\rangle^\mathrm{intra}
=
e\tau \sum_{n\mathbf{k}}
\bra{\psi_{n\mathbf{k}}}
\mathcal{O}
\ket{\psi_{n\mathbf{k}}}
\bra{\psi_{n\mathbf{k}}}
\boldsymbol{v}\cdot\mathbf{E}
\ket{\psi_{n\mathbf{k}}},
\end{equation}
and the interband response is driven by coherent superposition of Bloch states:
\begin{equation}
\left\langle 
\frac{d\mathcal{O}}{dt}
\right\rangle^\mathrm{inter}
=
-e\hbar \sum_{n\neq n'} \sum_\mathbf{k}
(f_{n\mathbf{k}} - f_{n'\mathbf{k}})
\mathrm{Im}
\left[ 
\frac{
\bra{\psi_{n\mathbf{k}}} 
d\mathcal{O}/dt
\ket{\psi_{n'\mathbf{k}}}
\bra{\psi_{n'\mathbf{k}}}
\boldsymbol{v}\cdot \mathbf{E}
\ket{\psi_{n\mathbf{k}}}
}{
\left( 
\mathcal{E}_{n\mathbf{k}} - \mathcal{E}_{n'\mathbf{k}} \right)^2 + \eta^2
}
\right],
\end{equation}
where $\psi_{n\mathbf{k}}$ is the Bloch state with band index $n$ and crystal momentum $\mathbf{k}$, $\mathcal{E}_{n\mathbf{k}}$ is its energy eigenvalue, $\boldsymbol{v}$ is the velocity operator, $\mathbf{E}$ is a spatially uniform dc electric field, $e>0$ is magnitude of the unit electric charge, $\hbar$ is the reduced Planck constant, and $\eta \rightarrow 0+$ is an infinitesimally small number. Detailed proof of Eq.~\eqref{eq:correspondence} can be found in Appendix A of Ref.~\cite{Go2020b}. 

In thin films of finite thickness, surface orbital accumulation can be quantified by \emph{orbital dipole}
\begin{equation}
\mathcal{P}_\alpha^{L_\gamma}
=
\frac{1}{2}
\left( 
r_\alpha L_\gamma + L_\gamma r_\alpha
\right),
\end{equation}
where $\alpha$ is the direction normal to the film's plane. By setting $\mathcal{O}=\mathcal{P}_\alpha^{L_\gamma}$ in Eq.~\eqref{eq:correspondence}, we obtain
\begin{equation}
\frac{1}{\tau}
\left\langle 
\mathcal{P}_\alpha^{L_\gamma}
\right\rangle^\mathrm{intra}
=
\left\langle 
\frac{d\mathcal{P}_\alpha^{L_\gamma}}{dt}
\right\rangle^\mathrm{inter},
\end{equation}
so the right-hand side is nothing but the definition of proper orbital current
\begin{equation}
\mathcal{J}_\alpha^{L_\gamma}
=
\frac{d\mathcal{P}_\alpha^{L_\gamma}}{dt}
=
j_\alpha^{L_\gamma} + \mathcal{P}_\alpha^{T_\gamma},
\end{equation}
where 
\begin{equation}
j_\alpha^{L_\gamma} = \frac{1}{2}
\left(
v_\alpha L_\gamma + L_\gamma v_\alpha
\right)
\end{equation}
is conventional current, and 
\begin{equation}
\mathcal{P}_\alpha^{T_\gamma}
=
\frac{1}{2}
\left( 
r_\alpha T_\gamma + T_\gamma r_\alpha
\right)
\end{equation}
is torque dipole with $T_\gamma = dL_\gamma /dt$. This proves the direct link between proper orbital current and orbital accumulation in thin films.

\begin{figure}[hb!]
\centering
\includegraphics[angle=0, width=0.8\textwidth]{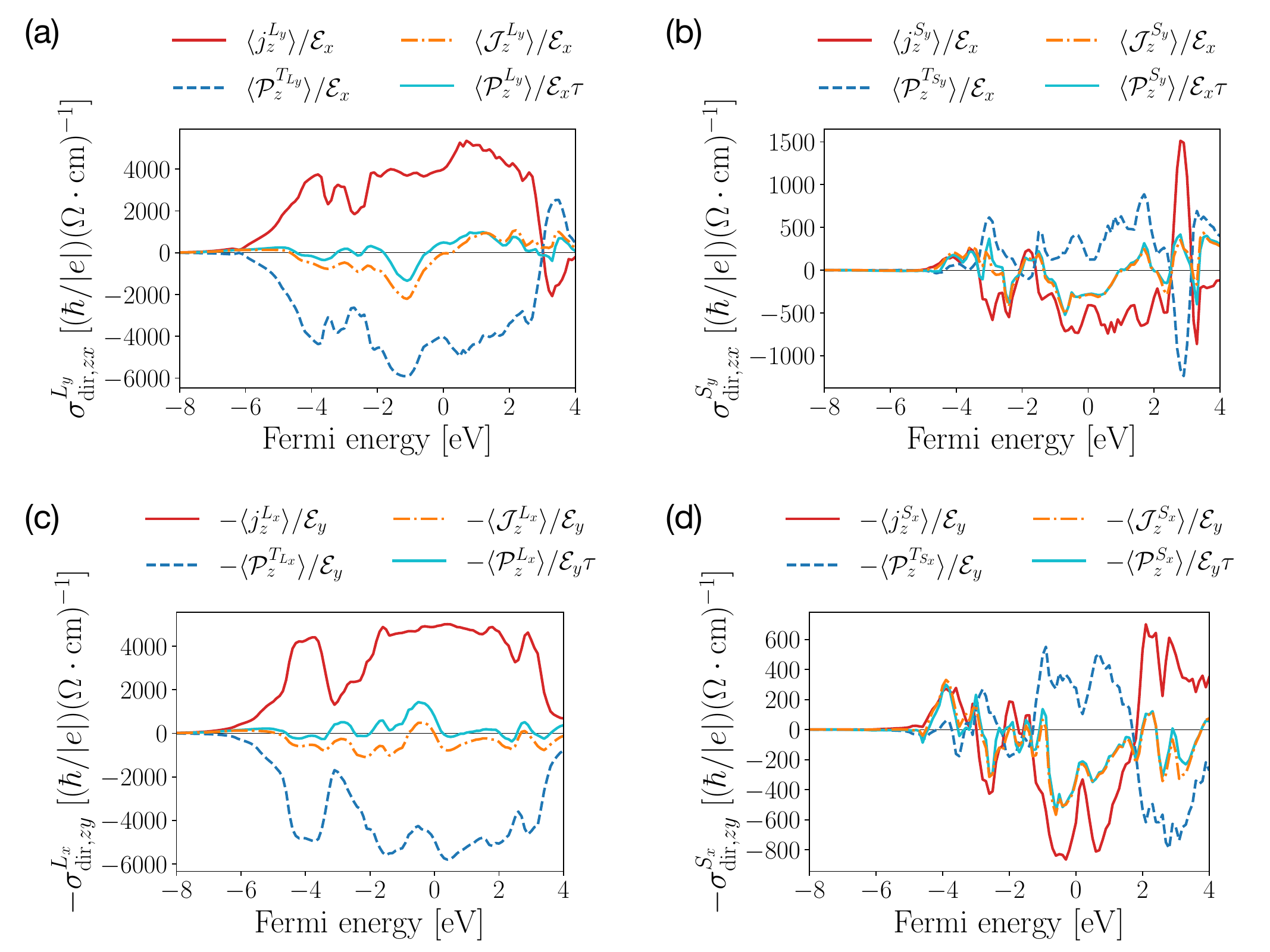}
\caption{
\label{fig:correspondence_W}
{Numerical demonstration of the correspondence between proper orbital/spin current and orbital/spin dipole in W(110).} The film has finite thickness of 33 atomic layers along bcc(110) direction, and the Cartesian coordinate is defined as $x\parallel [\bar{1}10]$, $y \parallel [001]$, and $z\parallel [110]$. Conventional current (red solid line), torque dipole (blue dahsed line), proper orbital/spin current (orange dash-dot line), and orbital/spin dipole (cyan solid line) are shown as a function of Fermi energy. (a) and (b) are for the orbital Hall conductivity $\sigma_{\mathrm{dir},zx}^{L_y}$ and the spin Hall conductivity $\sigma_{\mathrm{dir},zx}^{S_y}$, respectively, when an external electric field is applied along $x$. (c) and (d) show the results for $\sigma_{\mathrm{dir},zy}^{L_x}$ and $\sigma_{\mathrm{dir},zy}^{S_x}$, respectively, when an external electric field is applied along $y$.
}
\end{figure}

\begin{figure}[ht!]
\centering
\includegraphics[angle=0, width=0.8\textwidth]{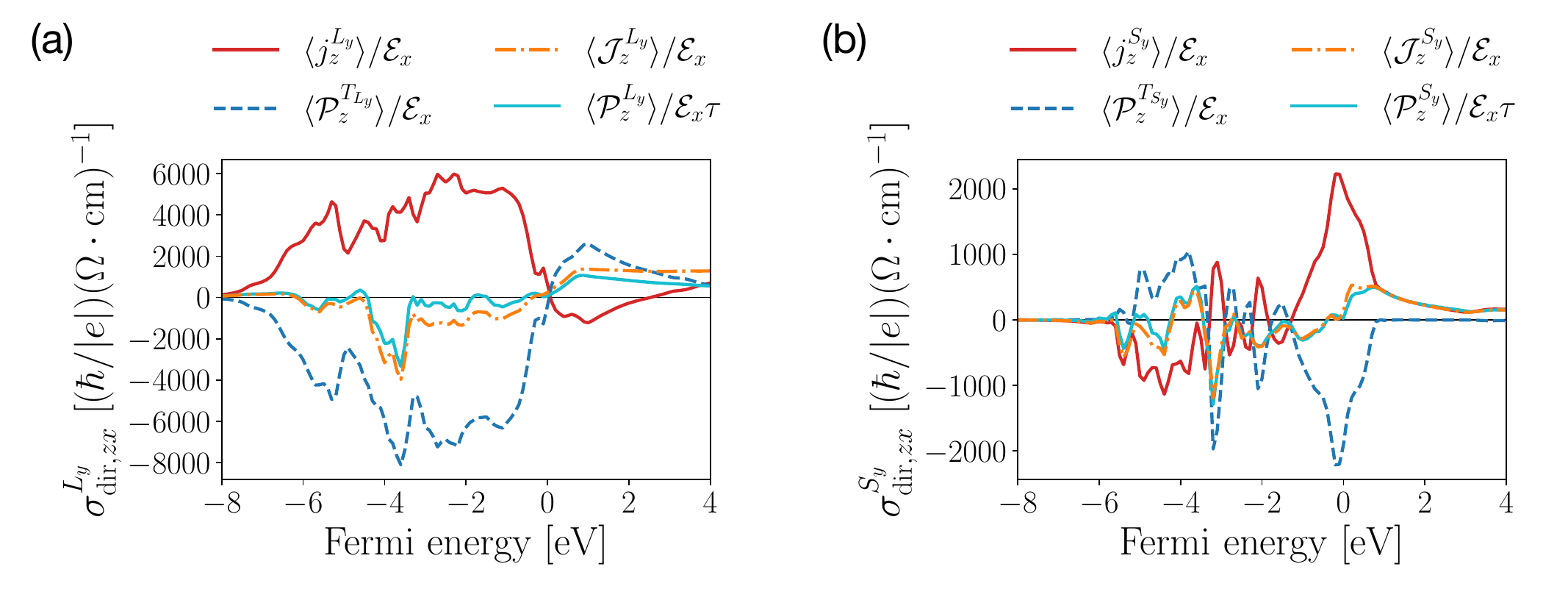}
\caption{
\label{fig:correspondence_Pt}
{Numerical demonstration of the correspondence between proper orbital/spin current and orbital/spin dipole in Pt(111).} The film has finite thickness of 33 atomic layers along fcc(111) direction, and the Cartesian coordinate is defined as $x\parallel [\bar{1}10]$, $y \parallel [\bar{1}\bar{1}2]$, and $z\parallel [111]$. Conventional current (red solid line), torque dipole (blue dahsed line), proper orbital/spin current (orange dash-dot line), and orbital/spin dipole (cyan solid line) are shown as a function of Fermi energy. (a) and (b) are for the orbital Hall conductivity $\sigma_{\mathrm{dir},zx}^{L_y}$ and the spin Hall conductivity $\sigma_{\mathrm{dir},zx}^{S_y}$, respectively, when an external electric field is applied along $x$. The results obtained when an external electric field is applied along $y$ are not shown because of weak anisotropy of the crystal structure.
}
\end{figure}

In Figs.~\ref{fig:correspondence_W} and \ref{fig:correspondence_Pt}, the interband responses of proper orbital/spin current and intraband responses of orbital/spin accumulation are shown as a function of Fermi energy for W(110) and Pt(111) films, respectively, both of which are composed of 33 atomic layers. The results for orbital and spin Hall effects are shown in (a) and (b), respectively, when an electric field is applied along $x$. For W(110) film [Fig.~\ref{fig:correspondence_W}], we also show the results when an external electric field is applied along $y$ in (c) and (d), which exhibit non-negligible difference compared to (a) and (b) due to anisotropy of the crystal structure. On the other hand, for Pt(111) [Fig.~\ref{fig:correspondence_Pt}], we show only the results when an electric field applied along $x$ because the anisotropy between $x$ and $y$ directions is negligible.

Over wide range of Fermi energy, the agreement between proper orbital/spin current (orange dash-dot lines) and orbital/spin dipole (cyan solid lines) is observed in both W(110) and Pt(111). While the agreement for the spin is nearly ideal, however, we find slight difference between proper orbital current and orbital dipole. This is because the orbital angular momentum operator acquires $\mathbf{k}$-dependence due to finite spread of Wannier functions, as also discussed in Appendix A of Ref.~\cite{Go2020b}. Another reason for the deviation is because Eq.~\eqref{eq:correspondence} is valid only in the limit of infinitesimally small $\eta$, but we assume $\eta=25\ \mathrm{meV}$ for improving the convergence of $\mathbf{k}$-space integral. 


\section{Additional Data on the Direct and Inverse Orbital/Spin Hall Effects}

\begin{figure}[h]
\centering
\includegraphics[angle=0, width=0.99\textwidth]{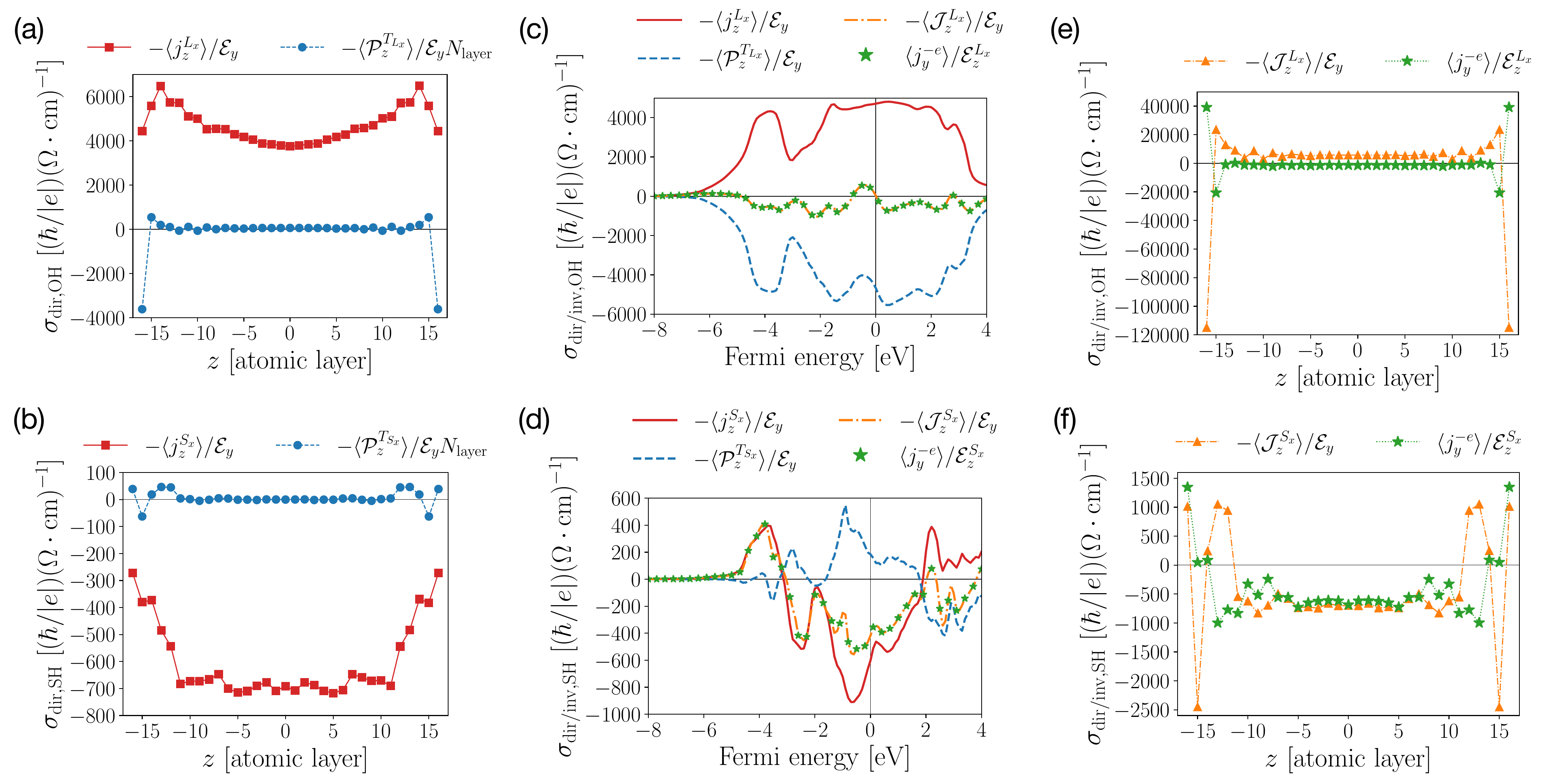}
\caption{
\label{fig:W_Ey}
First-principles calculation of the direct and inverse orbital Hall effects (a,c,e) and spin Hall effects (b,d,e) in a W(110) thin film for $L_x$ and $S_x$ polarization. (a,b) Local electric response of conventional orbital/spin current (red square symbols) and torque dipole (blue circle symbols) by the direct orbital/spin Hall effects. (c,d) Fermi energy dependence of the global electric responses of conventional orbital/spin current (red solid lines) and torque dipole (blue dashed lines). The sum of conventional orbital/spin current and torque dipole, proper orbital/spin current, is shown in orange dash-dot lines, which is exactly reciprocal to the response of charge current by orbital/spin voltage (green star symbols). (e,f) Comparison of the local responses of proper orbital/spin current (orange triangle symbols) and charge current (green star symbols) in the direct and inverse orbital/spin Hall effects, respectively. The local responses are substantially different in the direct and inverse orbital Hall effects, while those for the spin Hall effects are reciprocal even locally.
}
\end{figure}

\begin{figure}[h]
\centering
\includegraphics[angle=0, width=0.99\textwidth]{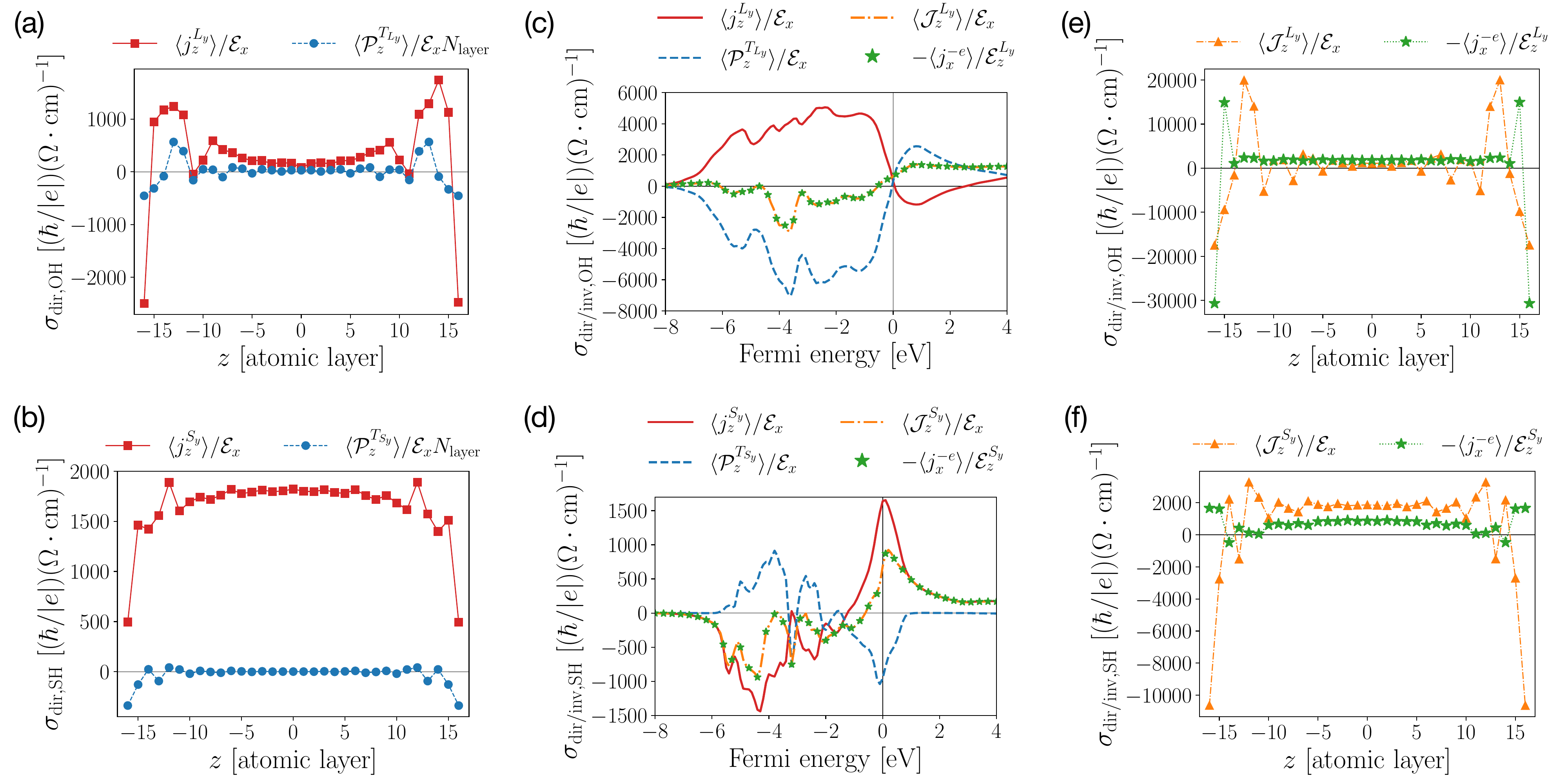}
\caption{
\label{fig:Pt_Ex}
First-principles calculation of the direct and inverse orbital Hall effects (a,c,e) and spin Hall effects (b,d,e) in a Pt(111) thin film for $L_y$ and $S_y$ polarization. (a,b) Local electric response of conventional orbital/spin current (red square symbols) and torque dipole (blue circle symbols) by the direct orbital/spin Hall effect. (c,d) Fermi energy dependence of the global electric responses of conventional orbital/spin current (red solid lines) and torque dipole (blue dashed lines). The sum of conventional orbital/spin current and torque dipole, proper orbital/spin current, is shown in orange dash-dot lines, which is exactly reciprocal to the response of charge current by orbital/spin voltage (green star symbols). (e,f) Comparison of the local responses of proper orbital/spin current (orange triangle symbols) and charge current (green star symbols) in the direct and inverse orbital/spin Hall effects, respectively. The local responses are substantially different in the direct and inverse orbital Hall effects, while those for the spin Hall effects are reciprocal even locally.
}
\end{figure}

In Tab. I of the main text, the numerical values of the response coefficients for the direct and inverse orbital/spin Hall effects are tabulated, but only the results for $L_y$ and $S_y$ polarization currents in W(110) are analyzed in Fig. 1. In Figs.~\ref{fig:W_Ey} and \ref{fig:Pt_Ex}, we show additional data for $L_x$ and $S_x$ currents in W(110) and for $L_y$ and $S_y$ currents in Pt(111), respectively. Note that the results for $L_x$ and $S_x$ currents are very similar to those for $L_x$ and $S_x$ currents, so we omit them here. All the data are obtained for $
\eta=0.1\ \mathrm{eV}$ while using the same computation parameters listed in Tab.~\ref{tab:params}.

\section{Decomposition into Conventional Current and Torque Dipole Contributions}

The orbital/spin Hall conductivities for the direct and inverse effects are decomposed into conventional current and torque dipole contributions. For the direct orbital Hall effect, the conventional current and torque dipole contributions are calculated locally in each layer at $z$ under a uniform electric field across the film:
\begin{subequations}
\begin{eqnarray}
\sigma_{\mathrm{OH,dir}}^{\mathrm{conv}}(z)
&=&
-\frac{e\hbar}{D}
\int \frac{d^2k}{(2\pi)^2}
\sum_{n\neq n'}
(f_{n\mathbf{k}}-f_{n'\mathbf{k}}) 
\sum_{z'}
\mathrm{Im}
\left[ 
\frac{
\bra{\psi_{n\mathbf{k}}}
j_z^{L_y} (z) 
\ket{\psi_{n'\mathbf{k}}}
\bra{\psi_{n'\mathbf{k}}}
v_x (z')
\ket{\psi_{n\mathbf{k}}}
}{(E_{n\mathbf{k}}-E_{n'\mathbf{k}})^2 + \eta^2}
\right],
\nonumber 
\\
\\
\sigma_{\mathrm{OH,dir}}^{\mathrm{torq}}(z)
&=&
-\frac{e\hbar}{D}
\int \frac{d^2k}{(2\pi)^2}
\sum_{n\neq n'}
(f_{n\mathbf{k}}-f_{n'\mathbf{k}}) 
\sum_{z'}
\mathrm{Im}
\left[ 
\frac{
\bra{\psi_{n\mathbf{k}}}
\mathcal{P}_z^{L_y} (z) 
\ket{\psi_{n'\mathbf{k}}}
\bra{\psi_{n'\mathbf{k}}}
v_x (z')
\ket{\psi_{n\mathbf{k}}}
}{(E_{n\mathbf{k}}-E_{n'\mathbf{k}})^2 + \eta^2}
\right],
\nonumber 
\\
\end{eqnarray}
\end{subequations}
where $D$ is the distance between neighboring atomic layers. On the other hand, in the inverse orbital Hall effect, perturbation is either conventional current and torque dipole, applied uniformly across the film, and the charge current response is computed for each atomic layer at $z$:
\begin{subequations}
\begin{eqnarray}
\sigma_{\mathrm{OH,inv}}^\mathrm{conv}(z)
&=&
\frac{e\hbar}{D}
\int \frac{d^2k}{(2\pi)^2}
\sum_{n\neq n'}
(f_{n\mathbf{k}}-f_{n'\mathbf{k}}) 
\sum_{z'}
\mathrm{Im}
\left[ 
\frac{
\bra{\psi_{n\mathbf{k}}}
v_x (z)
\ket{\psi_{n'\mathbf{k}}}
\bra{\psi_{n'\mathbf{k}}}
j_z^{L_y} (z')
\ket{\psi_{n\mathbf{k}}}
}{(E_{n\mathbf{k}}-E_{n'\mathbf{k}})^2 + \eta^2}
\right],
\nonumber 
\\
\\
\sigma_{\mathrm{OH,inv}}^\mathrm{torq}(z)
&=&
\frac{e\hbar}{D}
\int \frac{d^2k}{(2\pi)^2}
\sum_{n\neq n'}
(f_{n\mathbf{k}}-f_{n'\mathbf{k}}) 
\sum_{z'}
\mathrm{Im}
\left[ 
\frac{
\bra{\psi_{n\mathbf{k}}}
v_x (z)
\ket{\psi_{n'\mathbf{k}}}
\bra{\psi_{n'\mathbf{k}}}
\mathcal{P}_z^{L_y} (z')
\ket{\psi_{n\mathbf{k}}}
}{(E_{n\mathbf{k}}-E_{n'\mathbf{k}})^2 + \eta^2}
\right].
\nonumber 
\\
\end{eqnarray}
\end{subequations}
We remark that with the above definitions, the reciprocal relation is satisfied for the global volume-integrated quantities of each contribution,
\begin{subequations}
\begin{eqnarray}
\sum_z \sigma_{\mathrm{OH,dir}}^{\mathrm{conv}}(z)
=
\sum_z \sigma_{\mathrm{OH,inv}}^{\mathrm{conv}}(z),
\\
\sum_z \sigma_{\mathrm{OH,dir}}^{\mathrm{torq}}(z)
=
\sum_z \sigma_{\mathrm{OH,inv}}^{\mathrm{torq}}(z).
\end{eqnarray}
\end{subequations}
However, their spatially resolved quantities are generally different:
\begin{subequations}
\begin{eqnarray}
\sigma_{\mathrm{OH,dir}}^{\mathrm{conv}}(z)
\neq
\sigma_{\mathrm{OH,inv}}^{\mathrm{conv}}(z),
\\
\sigma_{\mathrm{OH,dir}}^{\mathrm{torq}}(z)
\neq
\sigma_{\mathrm{OH,inv}}^{\mathrm{torq}}(z).
\end{eqnarray}
\end{subequations}

\begin{figure}[ht!]
\centering
\includegraphics[angle=0, width=0.9\textwidth]{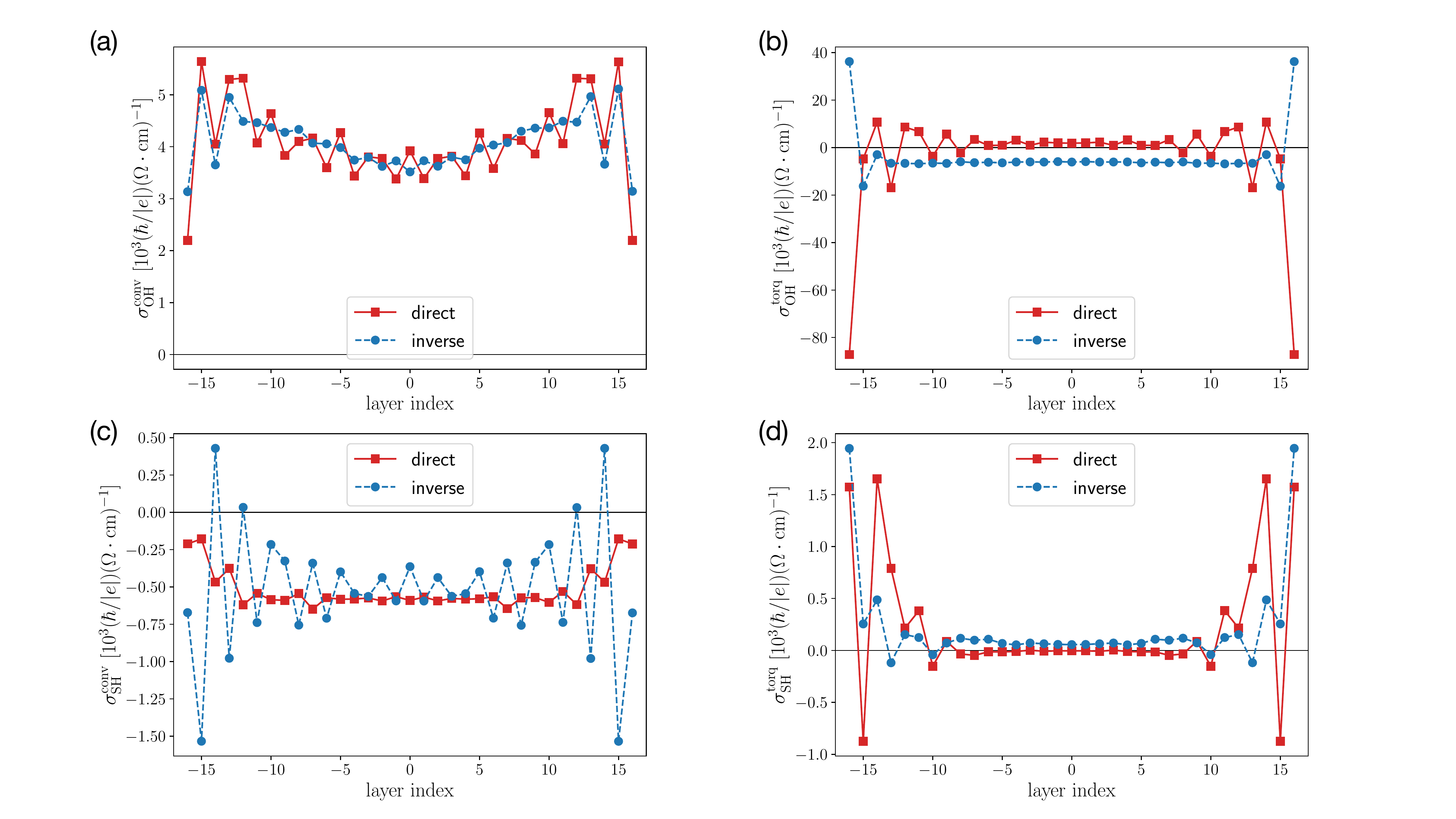}
\caption{
\label{fig:W_Ex_decompose}
Decomposition of the direct and inverse orbital Hall conductivities into (a) conventional current and (b) torque dipole contributions in W(110), for $L_y$ polarization and an electric field/current along $x$. (c) and (d) are conventional current and torque dipole contributions for the direct and inverse spin Hall conductivities, respectively.
}
\end{figure}

In Fig.~\ref{fig:W_Ex_decompose}, the results are shown for (a,b) $L_y$ and (c,d) $S_y$ currents in W(110). Figure~\ref{fig:W_Ex_decompose}(a) shows that the conventional current contributions for the direct and inverse orbital Hall effects are highly reciprocal even in their local distributions. However, the torque dipole contributions shown in Fig.~\ref{fig:W_Ex_decompose}(b) exhibit substantial contrast between the direct and inverse orbital Hall effects. The results for the spin Hall effects show the opposite behaviors: The conventional current contributions for the direct and inverse spin Hall effects differ in their spatial distributions, but the torque dipole contributions for the direct and inverse spin Hall effects are similar although they show significantly strong contributions at surfaces.

\begin{figure}[ht!]
\centering
\includegraphics[angle=0, width=0.9\textwidth]{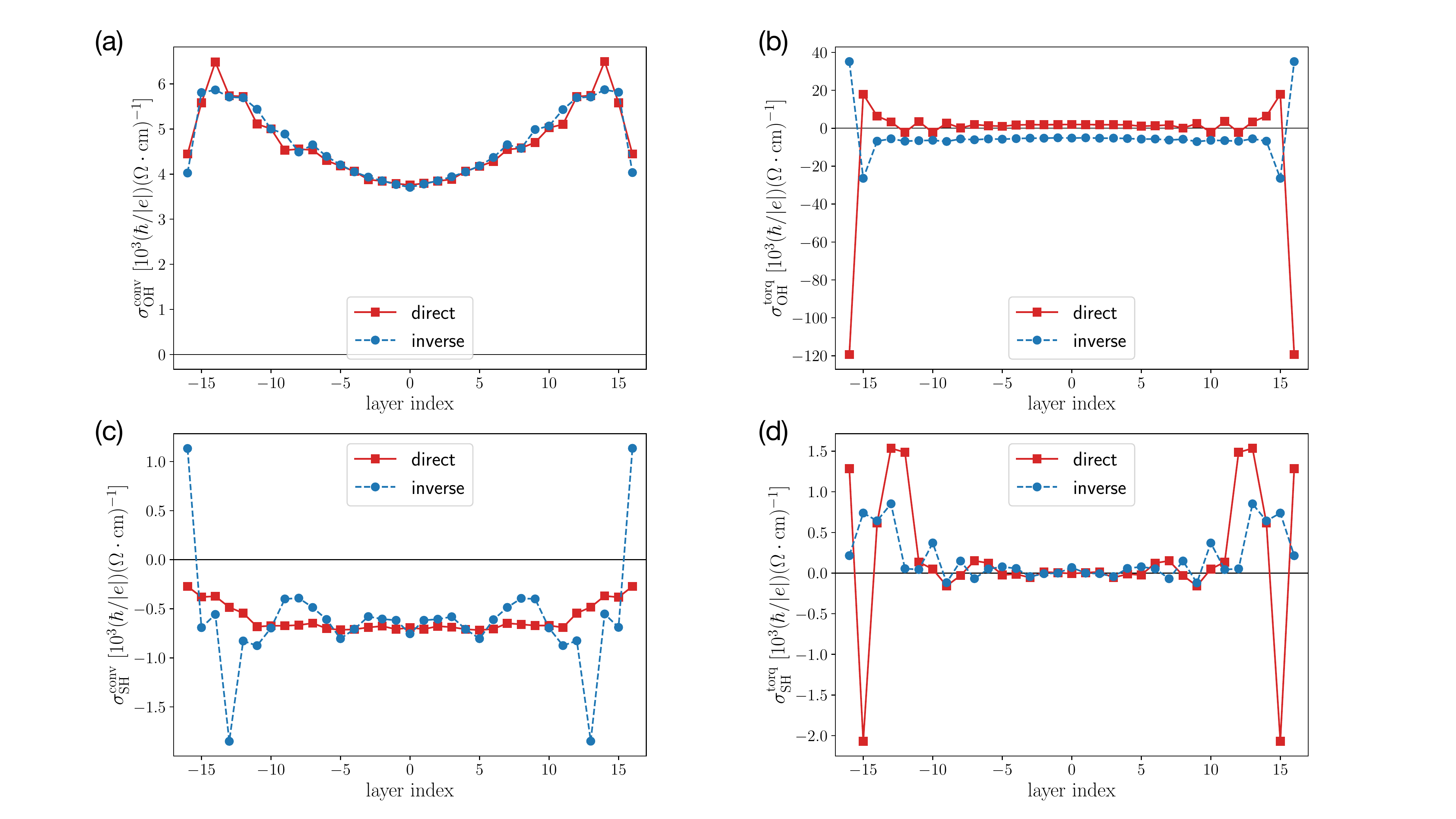}
\caption{
\label{fig:W_Ey_decompose}
Decomposition of the direct and inverse orbital Hall conductivities into (a) conventional current and (b) torque dipole contributions in W(110), for $L_x$ polarization and an electric field/current along $y$. (c) and (d) are conventional current and torque dipole contributions for the direct and inverse spin Hall conductivities, respectively.
\vspace{20pt}
}
\end{figure}

Because of the anisotropy of the structure of W(110), we also show the results for (a,b) $L_x$ and (c,d) $S_x$ currents in Fig.~\ref{fig:W_Ey_decompose}. General behaviors are similar to those for $L_y$ and (c,d) $S_y$ currents [Fig.~\ref{fig:W_Ex_decompose}], but the torque dipole contributions for the spin [Fig.~\ref{fig:W_Ey_decompose}(d)] exhibit significant difference between the direct and inverse effects at surfaces.

\begin{figure}[ht!]
\centering
\includegraphics[angle=0, width=0.9\textwidth]{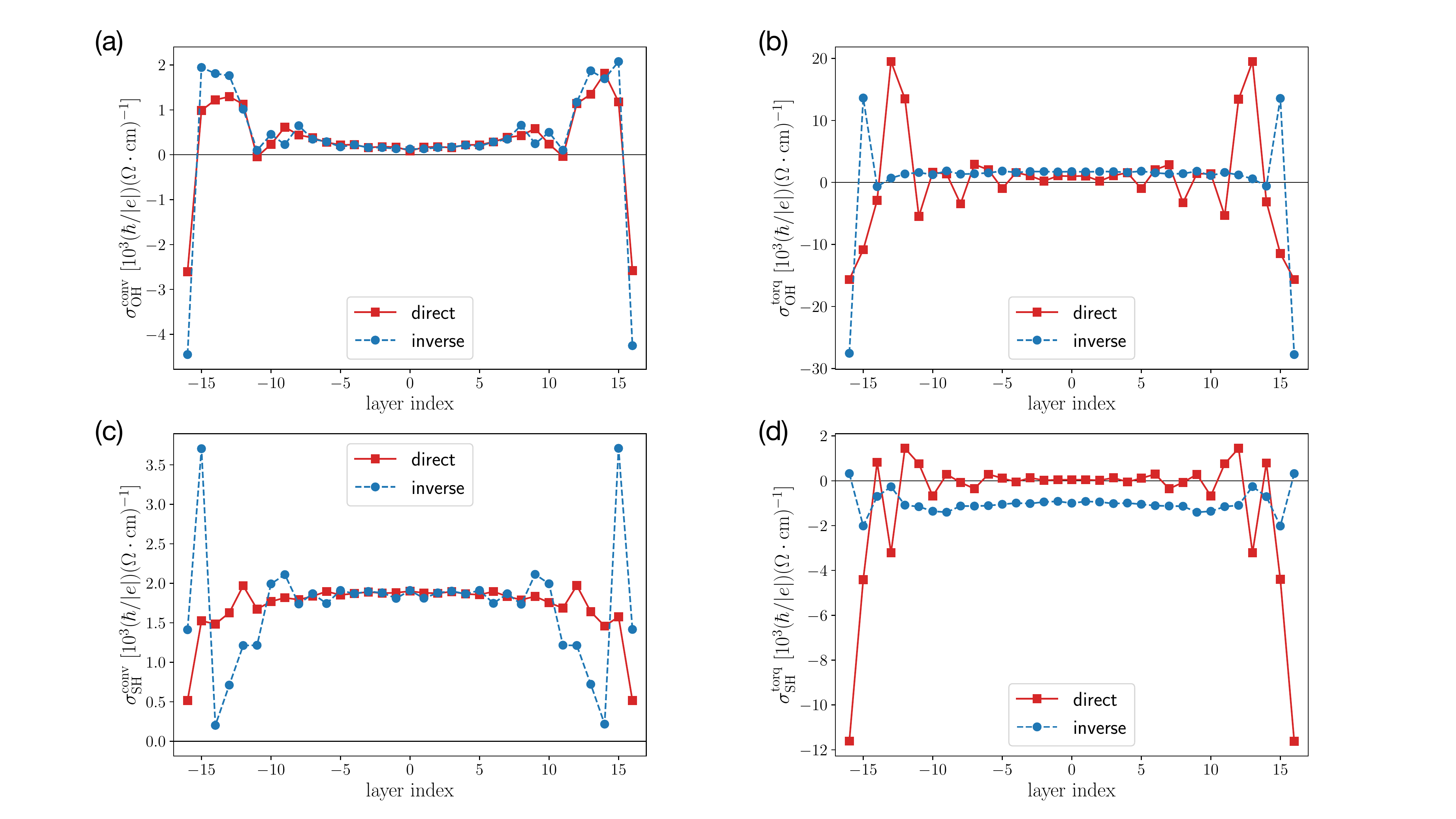}
\caption{
\label{fig:Pt_Ex_decompose}
Decomposition of the direct and inverse orbital Hall conductivities into (a) conventional current and (b) torque dipole contributions in Pt(111), for $L_y$ polarization and an electric field/current along $x$. (c) and (d) are conventional current and torque dipole contributions for the direct and inverse spin Hall conductivities, respectively.
}
\end{figure}

We show the results for Pt(111) in Fig.~\ref{fig:Pt_Ex_decompose}. For the orbital Hall effects, the conventional current contributions are reciprocal locally in each layer [Fig.~\ref{fig:Pt_Ex_decompose}(a)]. The torque dipole contributions exhibit substantial deviation at surfaces, but the bulk contributions are generally small [Fig.~\ref{fig:Pt_Ex_decompose}(b)]. The conventional current contributions in the spin Hall effects are reciprocal in the bulk, but the inverse spin Hall effect shows extremely high value at surfaces [Fig.~\ref{fig:Pt_Ex_decompose}(c)]. Interestingly, the torque dipole contributions for the spin Hall effects in Pt(111) [Fig.~\ref{fig:Pt_Ex_decompose}(d)] are highly nonreciprocal, similar to the torque dipole contributions to the orbital Hall effects in W(110) [Figs.~\ref{fig:W_Ex_decompose}(b) and \ref{fig:W_Ey_decompose}(b)]. We attribute this behavior to strong violation of the spin conservation due to large spin-orbit coupling in Pt.

\end{document}